\documentclass[lettersize,journal]{IEEEtran}

\PassOptionsToPackage{hyphens}{url}
\usepackage{setspace}
\usepackage{tabularx}
\usepackage{amsmath}
\usepackage{amsthm}
\usepackage{amssymb}    % 已包含 amsfonts 的功能

\usepackage{makecell}
\usepackage{longtable}
\usepackage{booktabs}
\usepackage{float}
\usepackage[table]{xcolor}   % 提前加载以支持表格颜色
\usepackage{multirow}
\usepackage{stfloats}

\usepackage[ruled,vlined]{algorithm2e}  % 选择 algorithm2e
\usepackage{graphicx}
\usepackage{subfig}
\usepackage{url}
\usepackage{cite}
\usepackage{textcomp}

\usepackage{flushend}
\usepackage[colorlinks,linkcolor=red,anchorcolor=green,citecolor=blue]{hyperref}
\usepackage{cleveref}

\usepackage{bbding}
\usepackage{verbatim}
\crefname{figure}{fig}{figures}
\Crefname{figure}{Fig}{Figures}
\begin{document}
\title{Agentic-SecPBFT: Agentic AI-Driven Proactive Security Framework for Wireless PBFT Consensus in Mobile Ad-Hoc Networks}

\author{Haoxiang Luo, Yinqiu Liu, Ruichen Zhang, Guangyuan Liu, Gang Sun,~\IEEEmembership{Senior Member,~IEEE},\\Hongfang Yu,~\IEEEmembership{Senior Member,~IEEE}, Zhu Han,~\IEEEmembership{Fellow,~IEEE}, and Dong In Kim,~\IEEEmembership{Life Fellow,~IEEE}

%\thanks{This work was supported by the Mobile Information Networks-National Science and Technology Major Project (2025ZD1302700, 2026ZD1307200).}

\thanks{H. Luo is with the WeBank-NTU Joint Research Institute on Fintech, Nanyang Technological University, Singapore 639798, and also with the College of Computing and Data Science, Nanyang Technological University, Singapore 639798 (e-mail:haoxiang.luo@ntu.edu.sg). Y. Liu, R. Zhang, and G. Liu are with the College of Computing and Data Science, Nanyang Technological University, Singapore 639798 (e-mail: yinqiu001@e.ntu.edu.sg; ruichen.zhang@ntu.edu.sg; liug0022@e.ntu.edu.sg.) G. Sun (corresponding author) and H. Yu are with the School of Information and Communication Engineering, University of Electronic Science and Technology of China, Chengdu 611731, China (e-mail: \{gangsun, yuhf\}@uestc.edu.cn). 
 Z. Han is with the Electrical and Computer Engineering Department, University of Houston, Houston, TX 77004, USA (email: hanzhu22@gmail.com). D. I. Kim is with the Department of Electrical and Computer Engineering, Sungkyunkwan University, Suwon 16419, South Korea (e-mail: dongin@skku.edu).}
}
 % <-this % stops a space
%\thanks{Manuscript received April 19, 2021; revised August 16, 2021.}}



\maketitle

\begin{abstract}
The standard Practical Byzantine Fault Tolerance (PBFT) protocol, designed for stable, wired environments, exhibits critical vulnerabilities when deployed in settings like mobile ad-hoc networks, thus making it susceptible to sophisticated threats such as Sybil attacks, Byzantine collusion, and message manipulation. Existing static defense mechanisms are ill-equipped to handle the intelligent and coordinated nature of these attacks. To address this challenge, this paper leverages the Agentic AI paradigm to build a distributed multi-agent system in which each consensus node is equipped with an intelligent agent. These agents employ a hierarchical Multi-Agent Deep Q-Network (MADQN) algorithm to learn and execute proactive security policies in real-time. By observing local network behavior, message consistency, and dynamically maintained reputation scores, the agents collaboratively identify suspicious behavior and recommend defensive actions under standard PBFT quorum and membership rules, thereby improving the integrity of the consensus process. We refer to the resulting framework as Agentic-SecPBFT. Then, we formally model key attack vectors and conduct extensive simulations. 
The results demonstrate that Agentic-SecPBFT reaches a 95.0\% attack detection rate with a 1.8\% false positive rate. Compared with mainstream PBFT variants, it achieves 3.1× higher throughput with 56\% lower latency on average under 33\% malicious nodes, offering a robust and adaptive security solution for decentralized wireless systems.

%The results demonstrate that, compared to other PBFT consensus, Agentic-SecPBFT achieves a high attack detection rate while significantly preserving network liveness and throughput, offering a robust and adaptive security solution for decentralized wireless systems.

\end{abstract}


\section{Introduction} \label{sec-I}
\subsection{Background}
\IEEEPARstart {T}{he} rapid proliferation of the Internet of Things (IoT) and the evolution of autonomous systems have catalyzed a fundamental architectural shift from centralized cloud computing to decentralized edge intelligence \cite{luo2025toward}. 
In this decentralized landscape, ensuring data integrity, traceability, and trust among inherently trustless entities is paramount \cite{luo2025wireless}. 
Blockchain technology, with its immutable ledger and distributed consensus mechanisms, has emerged as a foundational layer for these next-generation networks. Specifically, in wireless ad hoc networks such as Vehicular Ad Hoc Networks (VANETs) and Industrial IoT (IIoT) \cite{li2025blockchain}, \cite{chen2025blockchain}, blockchain enables secure Peer-to-Peer (P2P) transactions and trusted data sharing without reliance on a vulnerable central authority \cite{luo2024symbiotic}.   

However, the consensus mechanisms underpinning these blockchain systems, primarily the Practical Byzantine Fault Tolerance (PBFT) protocol, were originally conceived for static, high-bandwidth, and reliable wired networks, \cite{xu2022blown}. The direct transposition of PBFT to wireless mobile environments introduces critical performance bottlenecks and security vulnerabilities. Wireless links are inherently unreliable, subject to path loss, shadowing, and multipath fading phenomena such as Rician or Rayleigh fading \cite{luo2025convergence}. In a wired network, a timeout typically indicates a node failure or a network partition. In a wireless network, a timeout is frequently caused by deep fading or temporary interference. While standard PBFT interprets this as a malicious leader's silence, triggering expensive and often unnecessary view-change procedures that halt consensus progress and degrade throughput.   

Furthermore, the threat landscape in decentralized wireless networks has evolved significantly. Adversaries no longer rely solely on static, easily detectable malicious behaviors such as persistent packet dropping. Instead, they employ intelligent, adaptive strategies designed to exploit the limitations of traditional security protocols. For instance, selective dropping attacks \cite{gu2024dependable}, where malicious nodes oscillate between honest and dishonest behaviors to manipulate reputation scores. And Sybil attacks \cite{agrawal2025sybil}, where adversaries spawn multiple illegitimate identities to overwhelm voting processes, are particularly devastating in resource-constrained wireless networks. Existing solutions often rely on reactive reputation models or static thresholding. These methods fail to adapt to these complex, time-varying attack vectors, leading to a reactive security posture that is insufficient for critical infrastructure~\cite{li2021surveying}.

\subsection{Research Challenges}

Deploying a robust, secure, and efficient consensus mechanism in a wireless mobile environment presents a triad of primary technical challenges that this research aims to address:

\begin{itemize}
    \item \textbf{The Wireless Scalability-Reliability Trade-off:} Standard PBFT requires an $O(n^2)$ message complexity to reach consensus \cite{li2020scalable}, where $n$ represents the number of network nodes. In a wireless channel subject to interference, contention, and fading, the probability of successful message delivery decreases exponentially with network size and message volume \cite{yu2025distributed}. High packet loss rates trigger timeout mechanisms indistinguishable from malicious behavior, leading to view changes. Optimizing this trade-off requires a mechanism that can intelligently reduce message complexity without compromising Byzantine fault tolerance.
    
\item \textbf{Adaptive and Intelligent Adversarial Behavior:} Intelligent adversaries can exploit the forgetting factor of traditional trust models \cite{cheong2024multidimensional}. By behaving honestly to build a high reputation and then launching short-burst attacks, these nodes evade detection by static trust filters. Furthermore, in Sybil attacks, physically proximate nodes or a single node mimicking many can dominate the consensus group \cite{rafique2025trustworthy}. Distinguishing these attacks from legitimate behavior requires a system that learns temporal patterns and physical layer signatures.

\item \textbf{Convergence in Dynamic Topologies:} In Mobile Ad Hoc Networks (MANETs), the set of available validators changes rapidly due to node mobility. A consensus must distinguish between a node that has moved out of range. It belongs to a connectivity issue requiring a topology update. And a node that is withholding votes, which is a security issue requiring isolation \cite{xu2024puffchain}. Standard PBFT lacks the cross-layer awareness to make this distinction.
\end{itemize}

These challenges have been exacerbated by the rise of intelligent, coordinated attacks that bypass static defense mechanisms. For example, reputation-based PBFT variants  \cite{yang2025gaussian} can be deceived by attackers who build trust through honest behavior before launching attacks. Cryptographic solutions, such as Verifiable Random Functions (VRF) \cite{kang2024metaverses}, enhance primary node election security but fail to counter message manipulation attacks \cite{chen2025drdst}. 

The \emph{root of this vulnerability} lies in a semantic gap within the PBFT security model. Its security is built on a mathematical model that counts signed messages from $n$ distinct nodes to tolerate $f$ faulty ones. Here, $f$ represents the number of Byzantine nodes. In this model, a node is merely a cryptographic identity (i.e., a public key). However, in a wireless ad-hoc network, the real actors are physical devices. These devices are vulnerable to physical capture or software compromise and can easily generate multiple cryptographic identities, i.e., launch a Sybil attack \cite{benadla2022detecting}. Thus, an attacker controlling a single physical device can masquerade as multiple logical nodes, easily breaking the fundamental $n \geq 3f+1$  security assumption at the protocol level, unbeknownst to honest nodes. Static defense mechanisms, even reputation-based ones \cite{kang2024metaverses}, \cite{li2024reputation}, still operate at this flawed logical level, tracking the reputation of cryptographic keys rather than the trustworthiness of the underlying physical entities. Therefore, an effective defense system should reduce this semantic gap without treating a wireless fingerprint as an unforgeable credential. Our solution uses CSI/SINR observations and temporal behavior as probabilistic auxiliary evidence for associating logical identities with physical transmitters, while cryptographic signatures and PBFT quorum rules remain the basis of consensus correctness.

\subsection{Our Contributions}
To bridge this gap, this paper introduces Agentic-SecPBFT. 
A holistic security framework leverages the emerging Agentic AI paradigm \cite{zhang2025toward}, \cite{luo2026trustworthy} to transform consensus nodes from passive protocol executors into autonomous defenders. 
Unlike traditional designs, our framework employs a hierarchical Multi-Agent Deep Reinforcement Learning (MADRL) architecture. It comprises Local Consensus Agents (LCA) for real-time, device-level threat perception and a Cluster Optimization Agent (COA) for global strategy coordination. 
By utilizing a hierarchical Multi-Agent Deep Q-Network (MADQN) \cite{gronauer2022multi}, these agents construct dynamic behaviors of potential adversaries, enabling them to predict and preemptively isolate threats based on cross-layer behavioral patterns. To the best of our knowledge, this is the first attempt that uses agentic AI to enhance the wireless consensus security. The main contributions can be summarized as follows:

\begin{itemize}
    \item \textbf{Comprehensive Wireless and Attack Modeling:} We formulate a rigorous mathematical model that integrates wireless channel characteristics into the consensus failure probability. Furthermore, we formally define three distinct attack models, including Sybil, On-Off, and Oscillating Byzantine. It helps us to analyze their impact on network liveness and safety, establishing a baseline for evaluating defense mechanisms.   

  \item \textbf{Agentic AI-Enabled Security Framework:} We propose a hierarchical multi-agent architecture. LCA operates on individual nodes to evaluate peer behavior and channel states in real-time, while COA functions at a higher level to manage topology and global learning parameters. The COA is outside the PBFT commit-critical path. Its authenticated outputs guide learning and reputation estimation but cannot commit blocks, change quorum thresholds, or independently revoke membership.   

  \item \textbf{MADQN-Based Dynamic Consensus:} We develop a hierarchical MADQN algorithm that optimizes the selection of the primary node and the consensus group. The algorithm utilizes a novel reward function that balances throughput, latency, and security penalties. It allows the system to anticipate suspicious behavior and recommend proactive actions, while safety-critical exclusion and view changes remain subject to standard PBFT validation rules.

\end{itemize}

\subsection{Paper Structure}

The remainder of this paper is organized as follows. Section \ref{sec-II} provides a comprehensive review of related work in wireless consensus optimization and AI-driven security. Section \ref{sec-III} details the system model, including the network topology, channel models, and formal definitions of the attack vectors. Section \ref{sec-Iv} presents the Agentic AI framework, detailing the LCA and COA collaboration, the dynamic reputation mechanism, and the MADQN algorithm. Section \ref{sec-v} provides a security analysis of the framework against typical attacks. Section \ref{sec-vi} presents the extensive simulation results and discussion. Finally, Section \ref{sec-vii} concludes the paper.

\section{Related Works}\label{sec-II}

The optimization of consensus mechanisms for wireless and mobile environments is a rapidly evolving research area. Existing works can be broadly categorized into three streams, including communication complexity optimization, reputation-based trust mechanisms, and AI-driven security. In Table \ref{tab1}, we summarize the related works to highlight the innovations and differences of Agentic-SecPBFT.

\begin{table*}[!t]
\centering %
    \centering
    \caption{Related Works Comparison}
    
    % 设置线条宽度
    \renewcommand{\arrayrulewidth}{0.8pt} % 设置表格线宽度
    \renewcommand{\tabcolsep}{10pt} % 设置列间距
    
    {\fontsize{8}{10}\selectfont 
     
    \begin{tabular}{m{2.1cm}<{\centering}||m{1.3cm}<{\centering}|m{1.6cm}<{\centering}|m{1.4cm}<{\centering}|m{2.4cm}<{\centering}|m{1.7cm}<{\centering}|m{2.2cm}<{\centering}} % 定义每列宽度
        \hline
        \hline
         \textbf{Consensus} & \textbf{Intelligent Paradigm}  & \textbf{Reputation Management} & \textbf{Wireless Adaptation} & \textbf{Communication Complexity control} & \textbf{Active Defense} &\textbf{Semantic Gap Fill}\\ 
         \hline
         GRBFT \cite{ yang2025gaussian}  & \XSolidBrush & \Checkmark & \XSolidBrush & \Checkmark & \XSolidBrush & \XSolidBrush\\
        \hline
        DRDST  \cite{chen2025drdst} & \XSolidBrush & \Checkmark & \XSolidBrush & \Checkmark & \XSolidBrush & \XSolidBrush\\
        \hline
         RoUBC  \cite{wang2023robust} & \XSolidBrush & \XSolidBrush & \Checkmark & \Checkmark & \XSolidBrush & \XSolidBrush\\
        \hline
         WBFT \cite{luo2025weighted} & \XSolidBrush & \Checkmark & $\bigcirc$& \Checkmark & \XSolidBrush & \XSolidBrush\\
        \hline
        TRCO \cite{deng2024permissioned} & \XSolidBrush & \Checkmark & $\bigcirc$ & \Checkmark & \XSolidBrush & \XSolidBrush\\
        \hline
         DVRC \cite{zhang2024dynamic} & \XSolidBrush & \Checkmark & $\bigcirc$ & \Checkmark & \XSolidBrush & \XSolidBrush\\
        \hline
  DRC \cite{rashid2025trustworthy} & \XSolidBrush & \Checkmark & \XSolidBrush & \Checkmark & \XSolidBrush & \XSolidBrush\\
        \hline
         Xiong et al. \cite{xiong2025cloud}  & \XSolidBrush & \XSolidBrush & $\bigcirc$ & \Checkmark & \XSolidBrush & \XSolidBrush\\
        \hline
           ML BFT \cite{liu2026multi}  & \XSolidBrush & \XSolidBrush & \XSolidBrush & \Checkmark & \XSolidBrush & \XSolidBrush\\
        \hline
 
           DA-PBFT \cite{li2024dynamic}  & $\bigcirc$ & $\bigcirc$ & $\bigcirc$ & \Checkmark & \XSolidBrush & \XSolidBrush\\
        \hline
        Agentic-SecPBFT & \Checkmark & \Checkmark & \Checkmark & \Checkmark & \Checkmark & \Checkmark\\
        \hline
        \hline
    \end{tabular}}
 \vspace{2pt} % 微调和表格的间距
\footnotesize{$\bigcirc$: Partially mentioned; \Checkmark: Fully supported; \XSolidBrush: Not supported.}

    \label{tab1}
\end{table*}

\subsection{Communication Complexity Optimization in Wireless PBFT}

To mitigate the $O(n^2)$ message complexity in PBFT, which is prohibitive in bandwidth-constrained wireless networks, researchers have primarily focused on grouping strategies. The authors in \cite{chen2025drdst}, \cite{luo2024energy} discuss various grouping methods where the network is divided into clusters, and consensus is reached hierarchically. This grouping method, called sharding, can reduce communication costs by enabling multiple shards to reach consensus in parallel. However, it also introduces centralization risks at the shard heads \cite{fan2025shard}. Specifically, in high-mobility scenarios such as VANETs or Low-Altitude Wireless Networks (LAWNets) \cite{wang2023robust}, the overhead associated with frequent re-clustering and leader election often negates the benefits of reduced consensus traffic. Then it also leads to periods of instability where no consensus can be reached.

Additionally, recent work by models the wireless PBFT network using Stochastic Geometry and Poisson Point Processes (PPP) \cite{liu2025partially}. They derive the end-to-end success probability considering slotted ALOHA as the multiple access technique \cite{li2025timeliness}. Their framework allows for the calculation of optimal transmission intervals and viable coverage areas. While these provide a strong theoretical basis for physical layer optimization, it assumes honest nodes and does not account for Byzantine faults or intelligent jamming. Our work extends this by integrating the physical layer reliability model with an adversarial upper layer, bridging the gap between communication theory and consensus security.

\subsection{Reputation-Based Consensus Mechanisms}

Reputation systems aim to weigh votes or select leaders based on historical behavior, thereby excluding malicious nodes from the consensus process \cite{chen2025drdst}, \cite{luo2025weighted}. EigenTrust and similar models have been adapted for PBFT \cite{gao2019t}. For instance, Deng et al. \cite{deng2024permissioned}designed Trusted and Robust Consensus Optimization (TRCO) for VANETs, which can optimize the robustness and complexity of consensus. Meanwhile, Zhang et al. introduced a Dynamic Vehicle Reputation Consensus (DVRC) to enhance the data security and communication efficiency of the VANETs \cite{zhang2024dynamic}. These methods ensure that nodes with lower reputations will be excluded from the consensus group, or their voting weight will be reduced accordingly.

However, most existing reputation models use static decay factors. If a node behaves honestly for time $t$, its reputation $R$ increases linearly or exponentially \cite{chen2025drdst}, \cite{luo2025weighted}, \cite{rashid2025trustworthy}. Selective dropping attackers exploit this by calculating the exact time required to restore their reputation before attacking again. Furthermore, these systems typically fail to distinguish between packet loss due to poor channel conditions, such as fading, and malicious packet dropping, leading to high False Positive Rates (FPRs) in mobile environments. Therefore, we employ a dynamic, AI-driven reputation management. The Agentic AI analyzes the variance and volatility of a node's behavior.

\subsection{AI-driven Method in Consensus Security}

The application of AI, e.g., Reinforcement Learning (RL), to consensus security is a nascent but promising field. For instance, in \cite{altarawneh2025assessing}, the authors employed an AI-based attack detection method to verify that there were specific security vulnerabilities in the consensus, such as Denial of Service (DoS) attacks. Meanwhile, some work also proposes using RL to detect attacks in IoT networks by modeling the interaction as a Markov Decision Process (MDP)  \cite{ren2024blockchain}. Deep Q-Networks (DQN) \cite{kumar2024reinforcement} and Proximal Policy Optimization (PPO) \cite{xiong2025cloud} have been used to select validators to minimize latency and optimize throughput.

Current approaches largely utilize single-agent RL or centralized training with centralized execution. They suffer from scalability bottlenecks and create a single point of failure at the learning agent \cite{mohammadabadi2024communication}. Furthermore, few studies explicitly model the wireless channel state as part of the RL state space. A node might have low throughput due to deep fading, not malice; punishing such a node hurts network resilience and fairness. Thus, we utilize a MADRL approach with cooperative agents. We explicitly include Signal-to-Interference-plus-Noise Ratio (SINR) and Channel State Information (CSI) in the state space. This enables the agent to distinguish between channel errors and Byzantine faults, a critical distinction for mobile networks. Although there are a few works involving the MADRL and wireless consensus \cite{abishu2024multi}, they remain at their applications rather than focusing on the consensus security optimization.

\section{Network and Attack Models}\label{sec-III}
This section establishes the rigorous mathematical foundation for the proposed Agentic-SecPBFT framework. We explicitly model the stochastic nature of the wireless communications and formally define the adversarial capabilities that exploit the semantic gap between cryptographic identities and physical entities. We summary key notations in Table \ref{tab:notations}

\begin{table}[t!]
\centering
\caption{Summary of Notations}
\label{tab:notations}
    \renewcommand{\arrayrulewidth}{0.5pt} % 设置表格线宽度
    \renewcommand{\tabcolsep}{10pt} % 设置列间距
    
    {\fontsize{8}{10}\selectfont % 设置字体大小为8pt，行距为10pt
    \begin{tabular}{m{0.9cm}<{\centering}||m{6.5cm}} % 定义每列宽度
        \hline
        \hline
        \textbf{Symbol} & \textbf{Definition \& Physical Meaning} \\
        \hline
       % $CSI_{corr}$ & Spatial correlation matrix of Channel State Information for neighbor nodes \\
        $d_{ij}$ & Euclidean distance between node $i$ and node $j$ \\
     %   $Entropy_{vote}$ & Shannon entropy of the voting distribution of neighbor nodes \\
        $f$ & Number of Byzantine faulty nodes in the wireless network \\
        $f'$ & Perceived fault tolerance threshold in the network under Sybil attack \\
       % $F_{con}$ & Consensus-layer features for LCA real-time state perception \\
      %  $F_{net}$ & Network-layer features for LCA real-time state perception \\
      %  $F_{rep}$ & Reputation-layer features for LCA real-time state perception \\
        $h_{ij}$ & Small-scale fading coefficient modeling multipath effects \\
        $k$ & Number of physical nodes controlled by the adversary in the network \\
        $K_{Rice}$ & Rician $K$-factor, ratio of LoS to scattered power \\
        $N$ & Total number of physical nodes in the MANET network \\
        $N_0$ & Additive white Gaussian noise power spectral density \\
        $P_{rx,ij}$ & Received signal power at node $j$ from node $i$ \\
      %  $P_{success}(ij)$ & Expected message delivery probability from $i$ to $j$ (sigmoid of SINR) \\
        $P_{tx}$ & Transmission power of a wireless consensus node \\
        $R_{i,j}^t$ & Reputation score of node $j$ from node $i$ at time $t$ \\
        $SINR_{ij}$ & Signal-to-Interference-plus-Noise Ratio for link $i \to j$ \\
        $V_{avg}$ & Average movement velocity of the mobile nodes in MANET \\
        $\gamma_{th}$ & Minimum SINR threshold for successful packet decoding \\
        $\mu$ & Forgetting factor in dynamic reputation update \\
        $\Psi_{phy}$ & Physical certainty factor, weighting reputation penalty by channel quality \\
        $\sigma_f^2$ & Power of the scattered multipath components in Rician fading \\
        $\omega_{penalty}$ & Penalty weight for suspicious node behavior \\
        $\omega_{reward}$ & Reward weight for honest node behavior \\
        \hline
        \hline
    \end{tabular}}
\end{table}

\subsection{Wireless Consensus Network Model}
We consider a decentralized MANET where nodes communicate over an open, shared wireless medium. The network model incorporates both the mobility of physical devices and the stochastic characteristics of wireless channels, which are critical for distinguishing between benign failures and malicious behavior.

\emph{1) Node Mobility and Topology:}
The network consists of $N$ mobile nodes moving within a three-dimensional space $\mathbb{R}^3$. The Random Waypoint Model \cite{csahin2024reliable} governs the mobility of node $i$. In this model, a node selects a random destination within the simulation boundary and moves towards it with a velocity $v_i$ uniformly distributed in $[v_{min}, v_{max}]$. Upon reaching the destination, the node pauses for a duration $T_{pause}$ before selecting a new destination. This mobility creates a dynamic topology where the set of neighbors $\mathcal{N}_i(t)$ for any node $i$ changes continuously over time $t$.

\emph{2) Wireless Channel and Fading Model:}
Unlike wired networks, where link failures are binary, wireless links suffer from signal degradation \cite{xu2022blown}. We model the channel using a composite fading model that accounts for both large-scale path loss and small-scale multipath fading.

First, the received power $P_{rx,ij}$ is primarily determined by the distance $d_{ij}$. We employ a log-distance path loss mode:

 \begin{equation}
 P_{rx,ij}(dBm) = P_{tx}(dBm) - PL(d_0) - 10\eta \log_{10}\left(\frac{d_{ij}}{d_0}\right) + X_\sigma,
  \end{equation}
  where $d_0$ is the reference distance, $\eta$ is the path loss exponent, and $X_\sigma$ represents shadowing effects modeled as a zero-mean Gaussian random variable with standard deviation $\sigma$.
  
  Then, given the potential for LoS links in modern wireless applications, e.g., drone swarms or vehicular convoys, we utilize the Rician fading model \cite{tamura2024joint}. The channel coefficient $h_{ij}$ follows a Rician distribution, where the received signal amplitude $r$ has the Probability Density Function (PDF):

 \begin{equation}
 f(r) = \frac{r}{\sigma_f^2} \exp\left(-\frac{r^2 + \nu^2}{2\sigma_f^2}\right) I_0\left(\frac{r\nu}{\sigma_f^2}\right),
  \end{equation}
where $\nu^2$ is the power of the LoS component, $\sigma_f^2$ is the power of the scattered multipath components, and $I_0(\cdot)$ is the modified Bessel function of the first kind with order zero. The fading severity is characterized by the Rician K-factor, namely

\begin{equation}
K_{Rice} = \frac{\nu^2}{2\sigma^2}.
\end{equation}
A higher $K_{Rice}$ indicates a dominant LoS path and a more stable link, while $K_{Rice}=0$ reduces the model to Rayleigh fading, that is, non-LoS.

Next, the success of message reception is probabilistic. The instantaneous Signal-to-Interference-plus-Noise Ratio (SINR) at receiver $j$ is calculated as:

\begin{equation}
SINR_{ij} = \frac{P_{rx,ij} |h_{ij}|^2}{N_0 + \sum_{k \in \mathcal{T} \setminus \{i\}} P_{rx,kj} |h_{kj}|^2},
\end{equation}
where $\mathcal{T}$ is the set of simultaneously transmitting nodes. A consensus message $m$ is considered successfully received if and only if $SINR_{ij} \ge \gamma_{th}$ for the duration of the packet transmission. Here, $\gamma_{th}$ is the minimum required SINR at the receiver.

\subsection{Adversary Model and The Semantic Gap}

We define a robust Byzantine Adversary  with the following capabilities and constraints:   

\begin{itemize}
    \item \textbf{Physical Control:} The adversary fully controls $k$ physical nodes, where $k \le f$.
    
    \item \textbf{Cryptographic Agility:} The adversary can generate unlimited cryptographic key pairs ($pk, sk$) but cannot forge signatures of honest nodes.
    
    \item \textbf{Collusion:} All $k$ controlled nodes share a private out-of-band communication channel, allowing them to coordinate attacks instantly and share internal state.
    
    \item \textbf{Awareness:} The adversary is aware of the network topology and can estimate the reputation scores assigned to it by honest nodes.
    \item \textbf{Radio Frequency (RF) Manipulation:} The adversary may vary transmission power, mobility, and antenna behavior to influence channel observations, but cannot forge an honest node's digital signature or predict a fresh signed challenge before it is issued.
\end{itemize}

CSI is not assumed to be a cryptographic or inherently unforgeable identity. It is used only as probabilistic evidence together with fresh signed challenge--response results and temporally consistent observations from multiple LCAs. Let $\mathcal{E}_{phy}$ denote the event that active voting identities are correctly associated with distinct physical transmitters, and let
\begin{equation}
    \Pr[\mathcal{E}_{phy}] \geq 1-\epsilon_{phy},
\end{equation}
where $\epsilon_{phy}$ captures false splitting or false merging caused by fading, mobility, or deliberate RF manipulation. The security analysis explicitly conditions on $\mathcal{E}_{phy}$ and reports the corresponding residual probability rather than assuming perfect physical identification.

Moreover, the fundamental vulnerability we address is the semantic gap in classic consensus. In PBFT, it assumes a bijection between a consensus node and a physical entity (maybe a server). However, in MANETs, this bijection breaks down. The protocol sees $n$ public keys, while the physical reality contains only $N$ devices. If an attacker maps one physical device to $S$ public keys, the protocol operates under the illusion of high redundancy ($n \gg N$), while the physical resilience remains unchanged.

\subsection{Attack Models}
\begin{figure}[!t]
\centering
 \includegraphics[width=3.5 in]{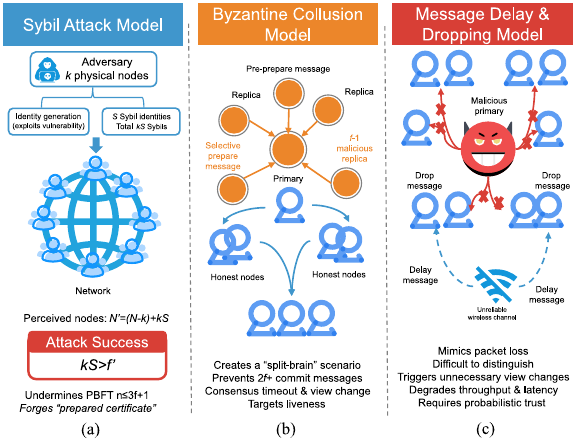}
   \caption{Attack models for wireless consensus networks. (a) Sybil attack; (b) Byzantine collusion; (c) Message delay and dropping.}
\label{fig:attack}
   \vspace{-0.5cm}
\end{figure}

The aforementioned semantic gap will give rise to the following three types of typical attacks, as shown in Fig. \ref{fig:attack}.

\subsubsection{Sybil Attack Model}

The Sybil attack is one of the most devastating attacks in wireless ad-hoc networks.
\begin{itemize}
    \item \textbf{Model}: Assume an adversary controls $k$ physical nodes ($k \le f$). By exploiting vulnerabilities in the identity creation process, the adversary generates $S$ Sybil identities for each controlled node. This changes the perceived total number of nodes in the network to $N' = (N-k) + k \cdot S$.
    \item \textbf{Mechanism}: This proliferation of identities directly undermines PBFT's $ n\geq 3f+1$ assumption. The condition for a successful attack can be formalized as
    \begin{equation}
        k \cdot S > f', \quad \text{where} \quad f' = \lfloor \frac{N'-1}{3} \rfloor.
    \end{equation}
    If the total number of Sybil identities controlled by the attacker $k \cdot S$ is greater than the new perceived fault tolerance threshold $f'$, the attacker can independently forge a “prepared certificate” for a malicious request. Thus, it can deceive honest nodes into believing the request has been endorsed by sufficient nodes.
    \item \textbf{Impact:} In MANETs, the dynamic topology of nodes further increases the difficulty in distinguishing Sybil identities from legitimate new nodes. The forged messages not only consume limited bandwidth and cause conflicts, but also make it difficult for honest nodes to collect sufficient valid votes. As a result, it triggers frequent view switches and paralyzes the consensus process \cite{wang2024scraft}.
\end{itemize}

\subsubsection{Byzantine Collusion Model (Selective Forwarding)}

This attack is more subtle than direct message forgery and uses node collusion to disrupt network liveness.
\begin{itemize}
    \item \textbf{Model}: Assume $f$ nodes are colluding, for instance, a malicious primary and $f-1$ malicious replicas.
    \item \textbf{Mechanism}: The malicious primary first broadcasts a valid \texttt{pre-prepare} message. In the subsequent \texttt{prepare} phase, the $f-1$ malicious replicas adopt a selective forwarding strategy: they send \texttt{prepare} messages to a subset of the honest nodes while remaining silent to the other half. Concurrently, these malicious nodes do not forward any messages from honest nodes, creating an information barrier between them.
   \item \textbf{Impact}: This leads to a ``split-brain” scenario. Some honest nodes may receive enough \texttt{prepare} messages and enter the `prepared` state, while others fail to do so due to insufficient messages. Ultimately, the network cannot gather $2f+1$ \texttt{commit} messages, causing a consensus timeout and triggering a costly view change process. This attack specifically targets system liveness, paralyzing the network by continuously creating deadlocks \cite{zheng2025trbft}.
\end{itemize}

\subsubsection{Message Delay and Dropping Model}

The most dangerous attacks in wireless networks are those that blur the line between inherent network instability (e.g., packet loss due to signal fading) and deliberate malicious behavior (e.g., selective dropping). A successful defense system must be able to probabilistically distinguish between these two conditions.
\begin{itemize}
    \item \textbf{Model}: A malicious primary aims to degrade system performance and liveness in an unnoticeable way.
    \item \textbf{Mechanism}: The primary can use the unreliability of the wireless channel as cover. It can selectively delay sending \texttt{pre-prepare} messages to specific honest nodes or drop them entirely. Since occasional packet loss is normal in a wireless environment, this behavior is difficult to immediately identify as malicious.
    \item \textbf{Impact}: The targeted nodes will time out and mistakenly assume the primary has failed, initiating a view change. The attacker can exploit this ambiguity to continuously trigger unnecessary leader elections, leading to a state of chronic disruption and inefficiency, severely reducing effective throughput and increasing transaction latency. A simple, threshold-based rule system is ineffective in this dynamic environment. It may misidentify honest nodes with poor link quality as malicious (false positives) or fail to identify attackers who cleverly manage their drop rates (false negatives) \cite{chao2024systematic}. %This indicates that trust decisions about nodes cannot be binary but must be probabilistic and context-dependent, which is precisely what reinforcement learning methods excel at. The agent can learn a policy that distinguishes environmental randomness from adversarial patterns.
\end{itemize}

\section{Agentic AI-Enabled Security Framework}\label{sec-Iv}
To counter these attacks, we employ Agentic AI to transform the consensus nodes from passive protocol executors to active and autonomous defenders. In this mode, Agentic AI's function is defined as follows:
\begin{itemize}
    \item \textbf{Contextual Perception:} Fusing upper-layer protocol states with physical-layer CSI to construct a holistic view of the adversarial environment.
\item \textbf{Strategic Autonomy:}
 Executing decentralized security policies that optimize long-term network liveness and integrity, rather than merely reacting to immediate protocol violations.

\item \textbf{Hierarchical Collaboration:} Leveraging a two-tiered multi-agent architecture to balance low-latency responses with strategic adaptation.
\end{itemize}

 %\begin{figure}[!t]
%\centering
% \includegraphics[width=3 in]{picture/agentic.pdf}
%   \caption{Agentic AI-enabled consensus nodes. They can autonomously perceive environmental information, including SINR, CSI, and node mobility, and formulate and execute attack defense strategies.}
%\label{fig:agentic}
%   \vspace{-0.5cm}
%\end{figure}

\subsection{Agentic Hierarchical Structure}

\begin{figure}[!t]
\centering
 \includegraphics[width=3.4 in]{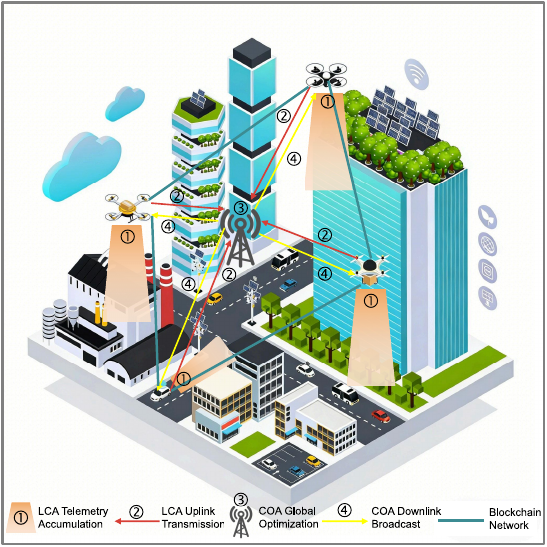}
   \caption{Hierarchical agent structure comprising LCAs and the COA. It consists of telemetry accumulation, uplink transmission, global optimization, and downlink broadcast.}
\label{fig:two}
   \vspace{-0.5cm}
\end{figure}

The complexity of securing a wireless consensus necessitates a separation of concerns. We define a hierarchical agent structure comprising LCAs and the COA, as shown in Fig. \ref{fig:two}. The framework employs a Centralized Training, Decentralized Execution (CTDE) paradigm, dividing responsibilities between two classes of agents.

\subsubsection{LCA Layer} The LCA is an embedded intelligent agent residing on every physical consensus node $i \in \mathcal{N}$. It is characterized by real-time execution and local observability. The LCA perceives the node's local environment state $s_t^{(i)}$, which includes SINR, CSI, etc. Also, the LCA is responsible for instantaneous threat detection and response. It executes the learned policy $\pi_{local}$ to make micro-decisions, such as verifying a suspect signature, challenging a potential Sybil node, or filtering messages from a low-reputation peer. Notably, operations must be completed within the PBFT phase timeout windows.

\subsubsection{COA Layer} Its primary role is not to make real-time security decisions but to guide the learning process of the individual LCAs. In a decentralized setting, the COA function is either hosted on a high-availability super-node, e.g., a base station in VANETs \cite{chen2025drdst}, or distributed via a Secure Multi-Party Computation (SMPC) committee \cite{li2025secure}. The COA aggregates telemetry from multiple LCAs to construct a quasi-global state representation $\mathcal{S}_{global}$. This allows it to detect patterns invisible to individual nodes, such as network-wide partition attacks or coordinated ``split-brain" voting strategies. Also, the COA is responsible for training the global policy network $\theta_{global}$ based on aggregated experiences and disseminating updated model parameters and global reputation scores back to the LCAs. Meanwhile, decisions occur on an epochal basis, e.g., every 100 blocks, allowing for computationally intensive optimization without stalling consensus.

\subsubsection{Interaction between COA and LCA} To enable robust collaboration between the  LCA and COA, we define the Agentic Interaction Protocol (AIP). The specific steps are as follows:

%The AIP is a standardized, asynchronous communication protocol designed to exchange security telemetry (uplink) and policy updates (downlink) with minimal overhead. It operates as an overlay on the existing blockchain P2P network. The AIP utilizes a compact binary message format to conserve wireless bandwidth. Interactions are defined by the tuple $\langle \text{Type}, \text{Payload}, \text{Sig} \rangle$, where message integrity is guaranteed by the node's Elliptic Curve Cryptography (ECC) signature \cite{zhou2025pmm}. 

\begin{itemize}
    \item \textbf{Step 1. Telemetry Accumulation:} During a consensus epoch $E$, the LCA buffers significant experience tuples, namely transitions where the reward prediction error $\delta > \epsilon$, into a local Experience Replay Buffer. Where $\delta=|Q(h_t,a)-Q_{target}(h_t,a)$, denoting the absolute error between the current Q-value and the target Q-value
    \item \textbf{Step 2. Uplink Transmission:} At the end of epoch $E$, the LCA compiles an \texttt{AIP\_TELEMETRY} packet and unicasts it to the COA.
    \item \textbf{Step 3. Global Optimization:} The COA aggregates received packets, performs a training step on the global MADQN, and computes the global reputation vector. 
    \item \textbf{Step 4. Downlink Broadcast:} The COA broadcasts an \texttt{AIP\_POLICY\_UPDATE} packet containing the new model weights and trust scores. The packet also carries the epoch number, monotonically increasing version, model digest, and COA authentication tag, enabling LCAs to verify integrity and reject replayed updates.
\end{itemize}

All AIP updates are signed, epoch-bound, and versioned. LCAs reject unauthenticated, stale, or rollback updates. Even a faulty COA cannot generate PBFT commit certificates, alter quorum thresholds, or permanently remove validators. If it is unavailable or detected as inconsistent, LCAs retain the latest authenticated policy and continue standard PBFT execution.

\subsection{Dynamic Trust and Reputation Management}

In wireless networks, a packet timeout can result from either deep fading (benign) or selective forwarding (malicious). Traditional models penalize both equally, leading to high false positives. We should define the expected delivery probability $P_{success}(i, j)$ based on the outage event of link $ij$ as follows:
%Then, we define the expected delivery probability $P_{success}(ij)$ based on the outage eventwe define the expected delivery: %probability $P_{success}(ij)$ as a sigmoid function of SINR:
\begin{equation}
P_{success}(i,j)=1-P_{out}(i,j),
\end{equation}
where $P_{out}(i,j)=\mathbb{P}[SINR_{ij} <\gamma_{th}]$ denotes the link outage probability. This should be evaluated by taking the expectations with respect to both the shadowing $X_{\sigma}$ in Eq. (1) and the fading $|h_{ij}|=r$ in Eq. (2).   %the minimum SINR for successful decoding and $\beta$ is a steepness factor.

Then, we model the reputation $R_{i,j}^t\in[0,1]$ as node $i$'s belief in node $j$'s honesty. To keep the score within its declared range while preserving the original incremental update, we use
\begin{equation}
\begin{aligned}
R_{i,j}^{t+1}=\Pi_{[0,1]}\!(R_{i,j}^{t}+(1-\mu)[\mathbb{I}_{honest}\omega_{reward}-\\\mathbb{I}_{suspect}\omega_{penalty}\Psi_{phy}]),
\end{aligned}
\end{equation}
where $\Pi_{[0,1]}(x)=\min\{1,\max\{0,x\}\}$ is the projection operator, $\mu$ is the forgetting factor, and $\mathbb{I}_{honest}$ and $\mathbb{I}_{suspect}$ indicate validated and suspicious behavior, respectively. The physical certainty factor is
\begin{equation}
\Psi_{phy}=P_{success}(i,j)(1-\xi_v),
\end{equation}
where $\xi_v=\min\{1,v_i/V_{max}\}$ attenuates confidence under high mobility. Thus, a timeout under a poor link causes only a small penalty, whereas a missing message under a reliable link causes a larger penalty. The reputation value is used for monitoring and node-selection recommendations; it cannot by itself revoke PBFT voting rights.

\subsection{Hierarchical MADQN for Proactive Defense}

The core decision engine of the LCA is the MADQN. Beyond proactive security defense, this framework also integrates agentic interference management to ensure the SINR in Eq. (4) meets the minimum rate requirement \(R_{\text{min}} = B \log_2(1+\gamma_{\text{th}})\) for consensus message transmission, where $B$ denotes the channel bandwidth. 
We formulate the security problem as a Decentralized Partially Observable Markov Decision Process (Dec-POMDP), defined by the tuple $\langle \mathcal{S}, \mathcal{A}, \mathcal{P}, \mathcal{O}, \mathcal{R}\rangle$. The goal is to learn a policy $\pi$ that maximizes the expected discounted return $J(\pi)=\mathbb{E}_{\pi}[\sum_{t=0}^{\infty}\gamma^t r_t]$.

\subsubsection{State Space $\mathcal{S}$}
The state space $\mathcal{S}$ is designed to capture the fingerprints of the three specific attack vectors. The observation vector $s_t^{(i)}$ for agent $i$ is a concatenation of consensus, network, and reputation features.

\begin{equation}
s_t^{(i)} = \left[ \mathbf{F}_{con}, \mathbf{F}_{net}, \mathbf{F}_{rep} \right],
\end{equation}
where $\mathbf{F}_{con}$ denotes the consensus features, including current PBFT phase $Phase_t$, view number $View_t$, the vector of recent voting behaviors of neighbors $\mathbf{V}_{history}$, and the shannon entropy of the voting distribution $Entropy_{vote}$, as shwon below. $\mathbf{F}_{net}$ shows the network features, including the vector of SINR values $\mathbf{\Gamma}_{vec}$ for neighbors $ SINR_{ij} $, the matrix of spatial correlation between CSI of different neighbors $\mathbf{CSI}_{corr}$, the variance in message arrival times $\Delta t_{arrival}$, 
the average interference intensity of the current slot $I_{\text{avg}}$, and the number of concurrent transmitters in the neighborhood $N_{\text{tx}}$.
$\mathbf{F}_{rep}$ is the reputation features, including current local reputation scores $\mathbf{R}_{local}$,variance of reputation over time $\sigma^2_{R}$.
\begin{equation}
Entropy_{vote}=-\sum_{v \in \{0,1\}}p(v)\log_2p(v),
\end{equation}
where $p(v)$ represents the probability of a neighbor voting in favor $1$ or against $0$. The higher the entropy value, the more chaotic the voting is, and the more likely there is a Byzantine collusion.

\subsubsection{Action Space $\mathcal{A}$}
The LCA has a discrete action space designed to execute specific security countermeasures. The action vector $a_t^{(i)}$ for agent $i$ includes:
\begin{itemize}
    \item \texttt{Participate Normally}: The default action when no threat is perceived.
    \item \texttt{Flag Suspicious Node}: A preliminary measure to increase monitoring of a specific node without immediate punitive action.
    \item \texttt{Decrease Reputation Score}: Locally adjust a neighbor's reputation based on evidence of misbehavior; a global reputation update must be delivered through an authenticated AIP update.
    \item \texttt{Request Challenge-Response}: Issue a cryptographic challenge to a suspicious node to verify its liveness and authenticity. This helps counter Sybil nodes that may be computationally constrained by forging multiple identities.
    \item \texttt{Vote for View Change}: Proactively initiate a view change when judging the primary's behavior to be malicious or inefficient.
    \item \texttt{Temporarily Reject Messages}: Temporarily deprioritize non-critical traffic from a node suspected of a severe attack (e.g., DoS flooding); valid signed PBFT votes remain governed by standard quorum and membership rules.
    \item \texttt{Adjust Transmission Power:} Dynamically tune transmission power within Upper and lower bounds of transmission power, namely $[P_{\text{tx,min}}, P_{\text{tx,max}}]$ to compensate for channel fading and mitigate interference.
  \item \texttt{Request Priority Slot:} Submit a reservation request to the COA for critical consensus messages to avoid collision with concurrent transmissions.
\end{itemize}

\subsubsection{Transition Probability $\mathcal{P}$}
Let $s_t=(c_t,p_t,r_t)$ denote the wireless-channel, PBFT-protocol, and reputation states, and let $a_t=(a_t^{phy},a_t^{sec})$ contain communication and security actions. Their coupled transition is factorized as
\begin{equation}
\begin{aligned}
\mathcal{P}(s_{t+1}|s_t,a_t)=&P(c_{t+1}|c_t,a_t^{phy})\\
&\cdot P(p_{t+1}|p_t,c_{t+1},a_t^{sec})\\
&\cdot P(r_{t+1}|r_t,p_{t+1},c_{t+1},a_t^{sec}).
\end{aligned}
\end{equation}
This factorization captures the dependence of consensus and reputation evolution on the realized wireless state, rather than treating channel and protocol dynamics as mutually exclusive alternatives.

\subsubsection{Observation Space $\mathcal{O}$}
Each LCA obtains a noisy local observation
\begin{equation}
o_{i,t}=\mathcal{O}_i(s_t,a_{t-1})+\varepsilon_{i,t},
\end{equation}
where the CSI/SINR components of $\varepsilon_{i,t}$ follow channel-estimation noise calibrated by link quality, while cryptographically verified protocol fields are treated as discrete observations. This avoids assuming one Gaussian model for all heterogeneous features.

\subsubsection{Reward Function $\mathcal{R}$}

The design of the reward function is critical for training effective defense agents. It is designed as a multi-objective function $r_t^{(i)}$ for agent $i$, calculated at each time step:
\begin{equation}
    r_t^{(i)} = w_1 \cdot r_{\text{consensus}} + w_2 \cdot r_{\text{security}} + w_3 \cdot r_{\text{liveness}}.
\end{equation}
\begin{itemize}
    \item $r_{\text{consensus}}$: A large positive reward for successfully committing a valid block; a large negative reward if consensus fails or a fraudulent block is committed.
    \item $r_{\text{security}}$: During simulation training, ground-truth attack labels provide positive rewards for confirmed detection and negative rewards for false alarms. During deployment or online adaptation, a security reward is generated only after verifiable evidence, such as an invalid signature, a failed fresh challenge--response, or a PBFT-certified view change; an unconfirmed local suspicion receives zero reward.
    \item $r_{\text{liveness}}$: A small negative reward for each time step that consensus stalls; a larger negative penalty for triggering a view change; a small positive reward for successfully maintaining \(SINR_{ij} \geq \gamma_{\text{th}}\) through power adjustment or reserving a priority slot for critical messages. This incentivizes the agent to maintain network progress, avoid unnecessary disruptive actions, and proactively optimize transmission conditions.
\end{itemize}
The weights $w_1$, $w_2$, and $w_3$ are tuned to prioritize security and correctness over raw performance. %The learned reward affects policy optimization only; PBFT certificate validity and membership changes remain governed by deterministic protocol rules.

 \begin{figure*}[!t]
\centering
 \includegraphics[width=6.5 in]{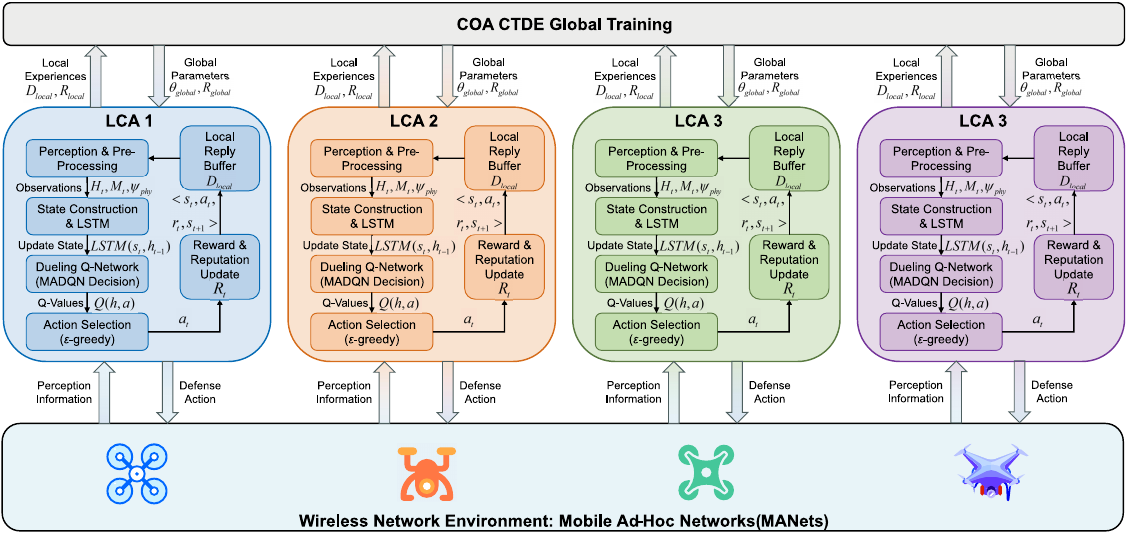}
   \caption{The algorithm steps of hierarchical MADQN, combining centralized training by COA and distributed execution by LCA.}
\label{CTDE}
   \vspace{-0.5cm}
\end{figure*}

\subsection{Algorithm Steps}
The algorithm on the LCA is shown as Fig. \ref{CTDE} and follows:

\begin{itemize}
    \item \textbf{Step 1. Perception and Pre-processing:} 
At time step $t$, the LCA collects raw physical layer metrics ($\mathbf{H}_t$) and protocol messages ($\mathbf{M}_t$). It computes the spatial correlation of CSI to detect potential Sybil clusters and calculates the physical certainty factor $\Psi_{phy}$.

   \item \textbf{Step 2. State Construction and Inference:}
The processed features form state $s_t$. This is fed into the local Q-network. To handle partial observability, we use an LSTM layer to aggregate historical states $h_t = \text{LSTM}(s_t, h_{t-1})$. The Dueling head then estimates the value:
\begin{equation}
Q(h_t, a) = V(h_t) + \left( A(h_t, a) - \frac{1}{|\mathcal{A}|}\sum_{a'} A(h_t, a') \right),
\end{equation}
where $V(h_t)$ represents the global state value. $A(h_t, a)$ is the action advantage value, which is output by an independent branch of the Dueling head. Both have the same dimension as the action space $\mathcal{A}$. The action $a_t$ is selected using an $\epsilon$-greedy policy.

\item \textbf{Step 3. Action Execution and Defense:}
If $a_t = \text{Challenge}(j)$, the LCA sends a fresh signed challenge. If $j$ fails or times out, $R_{i,j}$ is penalized according to the channel-aware rule. If $a_t = \text{Isolate}(j)$, only non-critical traffic is locally deprioritized; valid signed PBFT messages remain processable. Permanent exclusion requires a membership update approved by the underlying PBFT protocol. Otherwise, standard PBFT voting proceeds.

\item \textbf{Step 4. Reward Calculation and Storage:}
After the consensus round, the LCA evaluates the outcome (Block Committed vs. View Change). It calculates reward $r_t$ and stores the transition $\langle s_t, a_t, r_t, s_{t+1} \rangle$ in the local replay buffer.

\item \textbf{Step 5. CTDE Synchronization:}
Periodically, the LCA uploads samples to the COA via AIP. The COA trains the global model and returns updated weights, which the LCA integrates.

\end{itemize}

\begin{algorithm}[!t]
\caption{Centralized Training by COA for Agentic-SecPBFT}\label{alg:training}
\SetKwInOut{Input}{Input}
\SetKwInOut{Output}{Output}

\underline{Initialization:}\\
1. Initialize global Q-network parameters $\theta_{\text{global}}$ and target Q-network $\theta'_{\text{global}}$ \\
2. Initialize global replay buffer $\mathcal{D}_{\text{global}}$ and global reputation matrix $\mathbf{R}_{\text{global}} = \mathbf{0}$ \\
3. Broadcast initial $\theta_{\text{global}}$ and $\mathbf{R}_{\text{global}}$ to all LCAs \\
\BlankLine

\For{epoch = 1 to $E_{\text{max}}$}{
    \underline{Telemetry Aggregation:}\\
    1. Receive local experience buffers $\mathcal{D}_{\text{local}}^{(i)}$ and reputation matrices $\mathbf{R}_{\text{local}}^{(i)}$ from all LCAs $i$ \\
    2. Aggregate local experiences into $\mathcal{D}_{\text{global}}$ \\
    3. Compute global reputation $\mathbf{R}_{\text{global}}$ via weighted averaging of $\mathbf{R}_{\text{local}}^{(i)}$ \\
    \BlankLine
    
    \underline{Global Model Update:}\\
    1. Sample a batch of transitions from $\mathcal{D}_{\text{global}}$ if $D_{\text{global}} \geq B$ \\
    2. Calculate target Q-values $Q_{\text{target}} = r + \gamma \cdot \max_{a'} Q_{\theta'_{\text{global}}}(s', a')$ \\
    3. Update $\theta_{\text{global}}$ by minimizing MSE loss between $Q_{\text{target}}$ and $Q_{\theta_{\text{global}}}(s, a)$ \\
    4. Soft-update target network: $\theta'_{\text{global}} \leftarrow \tau \cdot \theta_{\text{global}} + (1-\tau) \cdot \theta'_{\text{global}}$ (every epochs) \\
    \BlankLine
    
    \underline{Parameter Broadcast:}\\
    1. Broadcast the signed and versioned update containing $\theta_{\text{global}}$ and $\mathbf{R}_{\text{global}}$ to all LCAs \\
    2. Clear $\mathcal{D}_{\text{global}}$ (or retain partial historical experiences) \\
}

\end{algorithm}

\begin{algorithm}[!t]
\caption{Decentralized Execution by LCA for Agentic-SecPBFT}\label{alg:lca_execution}
\underline{Initialization:}\\
1. Load global parameters $\theta_{\text{local}} \leftarrow \theta_{\text{global}}$ and $\mathbf{R}_{\text{local}} \leftarrow \mathbf{R}_{\text{global}}$ from COA \\
2. Initialize local replay buffer $\mathcal{D}_{\text{local}}$, $\epsilon$-greedy factor $\epsilon = \epsilon_{\text{init}}$, and LSTM hidden state $h_0 = \mathbf{0}$ \\
\BlankLine

\For{time step = 1 to $T_{\text{total}}$}{
    \underline{State Perception:}\\
    1. Extract consensus $\mathbf{F}_{\text{con}}$, network $\mathbf{F}_{\text{net}}$, and reputation $\mathbf{F}_{\text{rep}}$ features \\
    2. Construct local state $s_t = [\mathbf{F}_{\text{con}}, \mathbf{F}_{\text{net}}, \mathbf{F}_{\text{rep}}]$ and aggregate via LSTM: $h_t = \text{LSTM}(s_t, h_{t-1})$ \\
    \BlankLine
    
    \underline{Defense Action Selection:}\\
    1. Select action $a_t$ via $\epsilon$-greedy policy (random if rand() $<$ $\epsilon$, else $\arg\max_a Q_{\theta_{\text{local}}}(h_t, a)$) \\
    2. Execute $a_t$ under the PBFT safety guardrail (e.g., flag suspicious nodes, locally update reputation, or request a challenge) \\
    \BlankLine
    
    \underline{Reputation and Reward Update:}\\
    1. Update $\mathbf{R}_{\text{local}}$ based on neighbor behavior and channel quality ($\Psi_{\text{phy}}$) \\
    2. Calculate $r_t$, using ground-truth labels only in simulation and verified evidence during online adaptation \\
    3. Store transition $(s_t, a_t, r_t, s_{t+1})$ in $\mathcal{D}_{\text{local}}$ \\
    4. Decay $\epsilon$: $\epsilon \leftarrow \max(\epsilon_{\text{min}}, \epsilon \cdot \epsilon_{\text{decay}})$ \\
    \BlankLine
    
    \underline{COA Synchronization:}\\
    1. Upload $\mathcal{D}_{\text{local}}$ and $\mathbf{R}_{\text{local}}$ to COA (every $T_{\text{epoch}}$ steps) \\
    2. Receive and update $\theta_{\text{local}} \leftarrow \theta_{\text{global}}$ and $\mathbf{R}_{\text{local}} \leftarrow \mathbf{R}_{\text{global}}$ \\
}
\end{algorithm}

\subsection{Complexity Analysis}
The computational complexity of the MADQN-based Agentic-SecPBFT framework is split into two phases: offline training (COA) with pre-deployment cost, and online execution (LCA) for real-time edge performance.

\subsubsection{Offline Training Complexity (COA)}
The offline phase shown in Algorithm~\ref{alg:training}) is computationally intensive, with complexity dependent on max training epochs $E_{\text{max}}$, consensus rounds per epoch $T_{\text{epoch}}$, LCA number $N$, batch size $B$, and forward pass complexity of global Q-networks $C_Q^{\text{global}}$ and local Q-networks $C_Q^{\text{local}}$. Let $E_{\text{update}}$ denote parameter update steps per epoch.

\begin{enumerate}
    \item \textbf{Telemetry Aggregation:} All $N$ LCAs upload preprocessed local data to COA per $T_{\text{epoch}}$ round, with complexity $\mathcal{O}(T_{\text{epoch}} \cdot N \cdot C_Q^{\text{local}})$.
    
    \item \textbf{Global Model Update:} COA samples $B$ transitions and updates the global Q-network per $E_{\text{update}}$ step:
    \begin{itemize}
        \item Target Q-value calculation: $\mathcal{O}(B \cdot C_Q^{\text{global}})$;
        \item Network parameter update (MSE loss minimization): $\mathcal{O}(B \cdot C_Q^{\text{global}})$;
        \item Target network soft-update (every $C$ epochs): $\mathcal{O}(C_Q^{\text{global}})$, which is negligible.
    \end{itemize}
  % Then, the total update complexity is $
  %\mathcal{O}(E_{\text{update}} \cdot B \cdot C_Q^{\text{global}}).$
\end{enumerate}

Thus, the total COA training complexity is
$\mathcal{O}(E_{\text{max}} \cdot (T_{\text{epoch}} \cdot N \cdot C_Q^{\text{local}} + E_{\text{update}} \cdot B \cdot C_Q^{\text{global}}))$.

\subsubsection{Online Execution Complexity (LCA)}
The online phase shown in Algorithm~\ref{alg:lca_execution}) is lightweight, with complexity dependent on the total consensus rounds $T_{\text{total}}$, LCA number $N$, and neighbors per LCA $N_i(t)$.

\begin{enumerate}
    \item \textbf{State Perception \& Action Selection:} Each LCA constructs state $s_t^{(i)}$, runs LSTM aggregation, and selects actions via $\epsilon$-greedy policy per round: $\mathcal{O}(T_{\text{total}} \cdot |\mathcal{N}| \cdot C_Q^{\text{local}})$.
    
    \item \textbf{Reputation Update \& Synchronization:} 
    \begin{itemize}
        \item Reputation matrix update: $\mathcal{O}(N_i(t)|)$, it is negligible, due to $N_i(t) \ll N$);
        \item Reward calculation: $\mathcal{O}(1)$;
        \item COA synchronization (every $T_{\text{epoch}}$ rounds): negligible.
    \end{itemize}
\end{enumerate}

Thus, the total LCA execution complexity is $
\mathcal{O}(T_{\text{total}} \cdot N \cdot C_Q^{\text{local}})$.

\section{Security Analysis}\label{sec-v}
\subsection{Informal Analysis Against Attacks}
\subsubsection{Sybil Attack Defense}

Agentic-SecPBFT mitigates Sybil attacks by reducing the semantic gap between cryptographic identities and physical transmitters. Each LCA uses CSI spatial correlation and the Rician $K$-factor as probabilistic evidence rather than as an identity credential. Suspected co-located identities are further checked through fresh signed challenge--response exchanges and temporally consistent observations from multiple LCAs. The COA aggregates authenticated telemetry to identify network-wide Sybil patterns, and its result adjusts monitoring and reputation recommendations. Consequently, the framework reduces identity-proliferation risk with residual error $\epsilon_{phy}$.%; it does not claim that CSI alone guarantees a one-to-one mapping between keys and physical devices. %Under the RF manipulation model, deliberate fingerprint forgery can increase the physical association error $\epsilon_{phy}$, but cannot fully bypass the multi-evidence detection combining challenge-response and cross-LCA verification.

\subsubsection{Byzantine Collusion Defense}
Against selective forwarding collusion, LCAs monitor consensus-layer features, such as $Entropy_{vote}$, $V_{history}$, to detect ``split-brain" anomalies. The colluding nodes create inconsistent voting states, driving entropy beyond legitimate ranges. Then, LCAs execute defensive actions, e.g., reputation penalty, flagging, while the COA identifies coordinated patterns via global telemetry, updating $R_{global}$ to reflect systemic threats. Meanwhile, it reduces deadlocks caused by insufficient $2f+1$ commit messages through proactive monitoring, non-critical traffic deprioritization, and challenge--response, without discarding valid PBFT votes or changing the active membership directly. This can avoid reducing the system's activity due to passive view changes. Additionally, the hierarchical MADQN reward function strengthens early detection by rewarding threat mitigation and punishing consensus stalls.

\subsubsection{Message Delay/Dropping Defense}
The framework distinguishes benign fading from malicious message manipulation via the channel-aware metric $P_{success}$. LCAs also weight reputation penalties by $\Psi_{phy}=P_{success}$, including low SINR ($SINR_{ij} < \gamma_{th}$) attributes loss to fading (minimal penalty), while high SINR ($SINR_{ij} \gg \gamma_{th}$) infers malicious behavior (maximal penalty). Meanwhile, hierarchical MADQN’s state space integrates SINR vectors and message arrival variance, enabling LCAs to learn stealthy delay patterns. Additionally, the reward function minimizes false positives/negatives, ensuring honest nodes with poor channel conditions are not penalized. Attackers exploiting channel unreliability are flagged for additional verification and protocol-governed mitigation.

\subsection{Formal Security Analysis}

We separate PBFT correctness from the effectiveness of the learned defense policy. The following analysis does not assume that MADQN converges to an optimal policy, that LCA observations are error-free, or that the COA is always available. We use the following premises:

\begin{itemize}
    \item The active PBFT committee is configured with $N=3f+1$ voting identities, and at most $f$ corresponding physical validators are Byzantine.
    \item Digital signatures of honest validators cannot be forged, and an honest validator does not sign conflicting blocks at the same height and view.
    \item Learned actions cannot create commit certificates, change the $2f+1$ quorum, or permanently modify committee membership.
    \item The physical-association event $\mathcal{E}_{phy}$ occurs with probability at least $1-\epsilon_{phy}$; CSI is only auxiliary evidence for this event.
    \item AIP messages are authenticated and versioned. A faulty or unavailable COA may reduce detection quality, but its output remains advisory and is not required to complete PBFT.
\end{itemize}

\noindent \textbf{Theorem 1} (Policy-Independent Safety)
\emph{Conditioned on $\mathcal{E}_{phy}$, no two honest validators commit distinct blocks at the same height and view, regardless of the convergence state or detection errors of MADQN.}

\noindent \textit{Proof}: Assume that two honest validators commit conflicting blocks $B\neq B'$. Each block must contain a valid commit certificate with at least $2f+1$ distinct signatures. Let $\mathcal{Q}$ and $\mathcal{Q}'$ denote the two signer sets. Since $N=3f+1$,
\begin{equation}
|\mathcal{Q}\cap\mathcal{Q}'|\geq(2f+1)+(2f+1)-(3f+1)=f+1.
\end{equation}
At most $f$ physical validators are Byzantine under $\mathcal{E}_{phy}$; hence, the intersection contains at least one honest validator. This validator would have signed both $B$ and $B'$, contradicting the honest signing rule. MADQN and the COA cannot forge this signature or alter the quorum threshold. Therefore, safety is independent of the learned policy. Moreover,
\begin{equation}
\Pr[\mathsf{Safety}]\geq\Pr[\mathcal{E}_{phy}]\geq1-\epsilon_{phy}.
\end{equation}
$\hfill\blacksquare$

\noindent \textbf{Property 1} (COA Fault Containment)
\emph{A faulty COA can distribute a poor policy or biased reputation recommendation, but cannot by itself cause conflicting PBFT commits.}

\noindent \textit{Reasoning}: LCAs reject unauthenticated or stale AIP messages, while a valid COA update still contains no validator signatures for a block. Block commitment and permanent membership changes remain controlled by PBFT certificates. Thus, COA compromise affects detection efficiency and may increase latency, but does not bypass the safety quorum.

\noindent \textbf{Theorem 2} (Conditional Liveness)
\emph{Assume partial synchrony, eventual selection of an honest primary, and at least $2f+1$ responsive active validators after the network stabilization time. Then every valid request submitted to an honest validator is committed in finite time, independently of MADQN convergence and temporary COA unavailability.}

\begin{table}[t!]
\centering
\caption{Simulation Parameters}
\label{tab:sim_params}
{\footnotesize % Local font size setting
\begin{tabularx}{8.8cm}{ % Set total width
  >{\hsize=0.58\hsize\raggedright\arraybackslash}X  % Column 1: 55% width, left-aligned
  >{\hsize=0.42\hsize\raggedright\arraybackslash}X   % Column 2: 45% width, left-aligned
}
\hline\hline
\textbf{Parameter} & \textbf{Value} \\
\hline
\multicolumn{2}{c}{\textbf{Network Parameters}} \\
\hline
Simulation Area & 500 $\times$ 500 $\times$ 200 m \\
Number of Nodes ($N$) & 10-50 \\
%Mobility Model & Random Waypoint Model \\
Average Speed ($V_{\text{avg}}$) & 10-30 m/s \\
Pause Time ($T_{\text{pause}}$) & 1-5 s \\
Steepness Factor ($\beta$) & 2\\
\hline
\multicolumn{2}{c}{\textbf{Wireless Channel}} \\
\hline
Carrier Frequency & 2.4 GHz \\
Transmit Power ($P_{\text{tx}}$) & 1 W \\
Noise Power Density ($N_0$) & -174 dBm/Hz \\
SINR Threshold ($\gamma_{\text{th}}$) & 10 dB \\
Rician K-factor ($K_{\text{Rice}}$) & 2-10 dB \\
Transmission Power Range ($P_{\text{tx,min}}, P_{\text{tx,max}}$) & 0.1W, 2W \\
Channel Bandwidth ($B$) & 20 MHz \\
Minimum Rate Requirement ($R_{\text{min}}$) & 1 Mbps \\
\hline
\multicolumn{2}{c}{\textbf{Consensus \& PBFT}} \\
\hline
Block Size & 1024 bytes \\
Transaction Arrival Rate (TPS) & 10-30 \\
Malicious Node Ratio ($f/N$) & 0\% - 33\% \\
%View Change Timeout & Dynamically Adjusted \\
\hline
\multicolumn{2}{c}{\textbf{Dynamic Reputation Model}} \\
\hline
Forgetting Factor $\mu$ & 0.85 \\
Reward Weight $\omega_{reward}$ & 0.1 \\
Penalty Weight $\omega_{penalty}$& 0.5 \\
%View Change Timeout & Dynamically Adjusted \\
\hline
\multicolumn{2}{c}{\textbf{MARL Parameters}} \\
\hline
Learning Rate ($\alpha$) & 0.001 \\
Discount Factor ($\gamma$) & 0.99 \\
Epsilon ($\epsilon$-greedy) & Initial 1.0, decaying to 0.01 \\
$\omega_1, \omega_2, \omega_3$& 0.3, 0.5, 0.2\\
Experience Replay Buffer Size & 100,000 \\
Neural Network Architecture & 3-layer MLP (128, 256, 128) \\
\hline\hline
\end{tabularx}
}
\vspace{-0.5cm}
\end{table}

\noindent \textit{Proof}: After the network stabilization time, messages among responsive honest validators are delivered within a bounded delay. Once an honest primary is selected, it broadcasts a valid proposal and at least $2f+1$ responsive validators can complete the prepare and commit phases. Learned actions cannot lower the quorum, permanently exclude a validator, or suppress valid signed PBFT messages outside the standard protocol rules. If the COA is unavailable, LCAs continue with the latest authenticated policy and standard PBFT view change. Therefore, the system eventually forms a valid commit certificate. MADQN improves the probability of satisfying the reachability and responsiveness conditions by reducing false alarms and unnecessary view changes, but liveness does not rely on optimal learning convergence.
$\hfill\blacksquare$

\section{Performance Evaluation}\label{sec-vi}
\subsection{Simulation Setup and Baseline Schemes}

We build the simulation environment based on a time-synchronized joint framework of NS-3 and Python 3.9. Specifically, NS-3 performs high-fidelity modeling of wireless physical layer characteristics and MAC-layer transmission collisions, which provides realistic wireless channel states for the upper-layer consensus. The Python module runs the consensus, dynamic reputation management, and hierarchical MADQN algorithm. %The two modules interact via inter-process interfaces to synchronize network states and consensus events per time slot. 
It runs on a server equipped with three 96-core Intel(R) Xeon(R) Gold 5220R CPUs, 1 TB of memory, and 8 NVIDIA GeForce RTX 3090 GPUs. All parameters are listed in Table \ref{tab:sim_params}. 
To assess the convergence performance of the proposed Agentic AI-enabled hierarchical MADQN, we compare it with the advanced Double Deep Q-Network (DDQN) \cite{shang2025joint} and MADQN (without the COA in our hierarchical framework). Furthermore, to verify the optimization effect of this Agentic-SecPBFT, we also perform a performance comparison with the original PBFT, Gaussian Reputation-Based BFT (GRBFT) \cite{yang2025gaussian}, and Dynamic Adaptive PBFT (DA-PBFT) \cite{li2024dynamic}. Finally, we conduct a series of ablation experiments.

% \begin{figure}[!t]
%\centering
% \includegraphics[width=3 in]{picture/convergence1.pdf}
%   \caption{Convergence performance comparison. (a) $10\%$ malicious nodes; (b) $20\%$ malicious nodes.}
%\label{fig:C1}
%   \vspace{-0.5cm}
%\end{figure}

\subsection{Convergence Performance}
%Firstly, when the number of consensus nodes is $50$, we set the number of malicious nodes to $10\%$ and $20\%$, and compare the convergence performance of the proposed hierarchical MADQN and DDQN. The results are shown in Fig. \ref{fig:C1}. Whether with $10\%$ or $20\%$ malicious nodes, MADQN demonstrates significantly better convergence performance than DDQN within $2000$ training iterations. MADQN converges faster, with final average rewards remaining in the $130–170$ range and extremely small fluctuations. DDQN converges much slower, final average rewards only reaching $50-100$ and fluctuations $2-3$ times those of MADQN, with obvious oscillations and sudden reward drops in the $20\%$ malicious node scenario. The core reason for this gap lies in their adaptability to MANets. Hierarchical MADQN, based on a distributed two-layered architecture, leverages the autonomous decision-making and continuous learning of the Agentic AI paradigm. It allows LCAs to real-time monitor neighboring nodes’ RTT and packet loss rates and proactively optimize consensus to mitigate malicious node impacts. In contrast, DDQN lacks a collaborative mechanism, and its single agent struggles to cope with the network’s dynamic topology, unstable channels, and erroneous feedback from malicious nodes.

 \begin{figure*}[!t]
\centering
 \includegraphics[width=6.5 in]{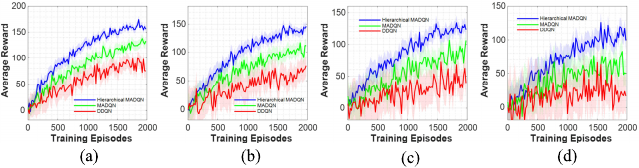}
   \caption{Convergence performance comparison. (a) Low mobility \& low interference; (b) High mobility \& low interference; (c) Low mobility \& high interference; (d) High mobility \& high interference.}
\label{fig:C2}
   \vspace{-0.5cm}
\end{figure*}

%Secondly, 
We compare the convergence performance in four combinations of low mobility ($V_{avg}=10$ m/s) and high mobility ($V_{avg}=30$ m/s), high interference ($T_{pause}=1$ s) and low interference ($T_{pause}=5$ s). 
The interference here refers to the communication instability caused by node mobility, dynamic topology changes, and wireless channel fading. Across all four scenarios, hierarchical MADQN achieves the best convergence performance, followed by  MADQN, while DDQN consistently lags in both convergence speed and stability.
In Fig. \ref{fig:C2} (a), hierarchical MADQN converges rapidly, with final average rewards maintained in the 160–180 range and slight fluctuations. MADQN, lacking global COA coordination, converges at a moderate pace and stabilizes around 130–150, outperforming DDQN but falling short of the hierarchical design. DDQN converges more slowly and has fluctuations approximately twice that of hierarchical MADQN.
In Fig. \ref{fig:C2} (b), hierarchical MADQN still maintains stable convergence, with rewards remaining in the 140–160 range. MADQN shows more noticeable oscillations under high mobility but still retains a clear performance advantage over DDQN. In contrast, DDQN exhibits more obvious oscillations and occasional reward drops due to high node mobility.
For Fig. \ref{fig:C2} (c) and Fig. \ref{fig:C2} (d), hierarchical MADQN’s advantages are also prominent. It can quickly adapt to the unstable communications, with final average rewards remaining at 130-150 and 110-130, respectively, as the environment becomes more challenging. MADQN follows with moderate performance degradation, while DDQN struggles to cope with the combined impact of interference and mobility, showing slow convergence speed and severe reward oscillations in the high mobility and high interference scenario.
The reason is that in hierarchical MADQN, LCAs can perceive channel states and node mobility changes in real time, while the COA optimizes global strategies based on aggregated telemetry. It enables the system to proactively adapt to environmental changes. %MADQN retains distributed local perception but lacks global strategy guidance, resulting in limited adaptation capability under dynamic environments. In contrast, DDQN cannot accurately distinguish between environmental interference and mobility-induced topology changes, leading to erroneous learning feedback and poor convergence stability.

\subsection{Consensus Security}
We randomly run 50-round simulations based on Table \ref{tab:sim_params}, to compare the consensus security under attack conditions. Specifically, we simulate under different simulation steps, malicious node ratios, and interference intensity.

\begin{figure*}[!t]
\centering
 \includegraphics[width=6.5 in]{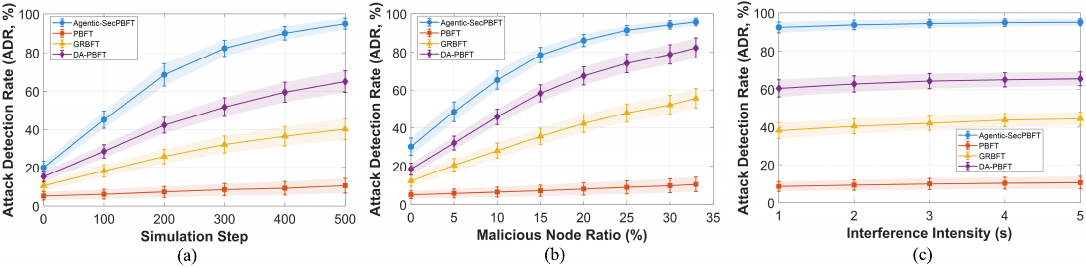}
   \caption{Attack detection rate comparison. (a) Under different simulation steps; (b) Under different malicious node ratios; (c) Under different interference intensities.}
\label{fig:ADR}
   \vspace{-0.5cm}
\end{figure*}

The three Attack Detection Rate (ADR) comparison results shown in Fig. \ref{fig:ADR} highlight Agentic-SecPBFT’s unique proactive defense capability. Fig. \ref{fig:ADR} (a) shows that its ADR climbs to $95.0\%$ after 500 steps, benefiting from the hierarchical MADQN algorithm’s continuous learning of adversarial behavioral patterns. In contrast, PBFT’s ADR remains below $11\%$ without a detection mechanism. GRBFT and DA-PBFT’s rates stall at $40.2\%$ and $65.0\%$ respectively. Fig. \ref{fig:ADR} (b) reveals that Agentic-SecPBFT’s ADR increases with the attack intensity, reaching $95.8\%$ at $33\%$ malicious nodes. This is because more attack samples enhance the hierarchical MADQN’s pattern recognition accuracy. Fig. \ref{fig:ADR} (c) demonstrates its anti-interference advantage. Even at the highest interference, Agentic-SecPBFT's ADR remains $92.5\%$, while GRBFT and DA-PBFT’s rates drop to $38.2\%$ and $60.5\%$, respectively. Agentic-SecPBFT’s excellent attack detection performance is attributed to its Dec-POMDP formulation. It integrates consensus-layer features, network-layer features, and reputation-layer features into a unified state space, making the hierarchical MADQN algorithm capable of capturing the multi-dimensional fingerprints of attacks. Also, the COA aggregates global telemetry data from all LCAs, enabling the identification of network-wide coordinated attack patterns that are invisible to a single node. Additionally, it fuses physical-layer CSI/SINR information to distinguish malicious behavior from channel fading caused by interference, thus maintaining high detection accuracy in complex wireless environments.

\begin{figure*}[!t]
\centering
 \includegraphics[width=6.5 in]{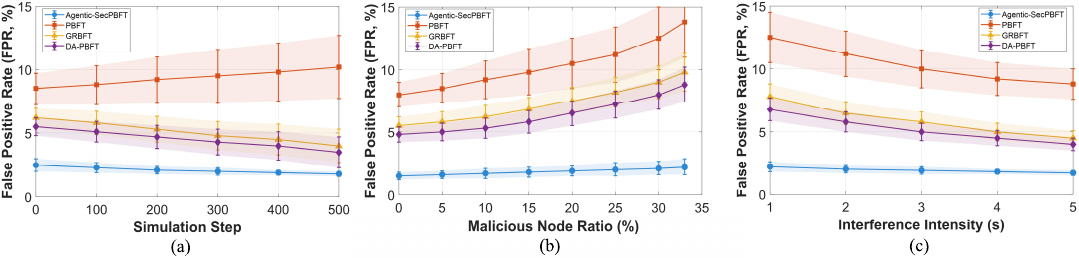}
   \caption{False positive rate comparison. (a) Under different simulation steps; (b) Under different malicious node ratios; (c) Under different interference intensities.}
\label{fig:FPR}
   \vspace{-0.5cm}
\end{figure*}

Moreover, Agentic-SecPBFT’s high detection precision is reflected in its consistently low False Positive Rate (FPR) across all scenarios. Fig. \ref{fig:FPR} (a) shows its FPR stabilizes at $1.8\%$ after 500 steps, far below PBFT’s $10.2\%$, GRBFT’s $4.0\%$, and DA-PBFT’s $3.5\%$. Fig. \ref{fig:FPR} (b) indicates that even with $33\%$ malicious nodes, Agentic-SecPBFT’s FPR remains at $2.2\%$, as its dynamic reputation mechanism filters out false alarms. Fig. \ref{fig:FPR} (c) confirms its ability to avoid misjudging interference as malicious. when $T_{pause}$ decreases from $5$s to $1$s, its FPR only rises from $1.7\%$ to $2.2\%$, while PBFT’s FPR surges from $8.8\%$ to $12.5\%$. The higher FPR of others is due to their lack of physical layer context awareness, leading to indiscriminate penalties for packet loss or node unresponsiveness. The key to Agentic-SecPBFT’s ultra-low FPR is its channel-aware reliability metric $P_{success}$. It quantifies the message delivery probability based on real-time channel conditions. And the physical certainty factor $\Psi_{phy}$ weights the reputation penalty for suspicious behavior. It weakens the penalty when the channel quality is poor to avoid misjudging honest nodes affected by fading, and imposes maximum penalties only when the channel is reliable. In addition, its dynamic reputation update rule combines a forgetting factor with channel quality. It can effectively filter environmental noise and ensure that honest nodes with poor channel conditions are not misclassified.%, thus maintaining network fairness.

\subsection{Consensus Performance}
Then, we compare their consensus success rate, latency, and throughput under different conditions, such as the malicious node ratio, the number of nodes, and the speed of nodes.

\begin{figure*}[!t]
\centering
 \includegraphics[width=6.5 in]{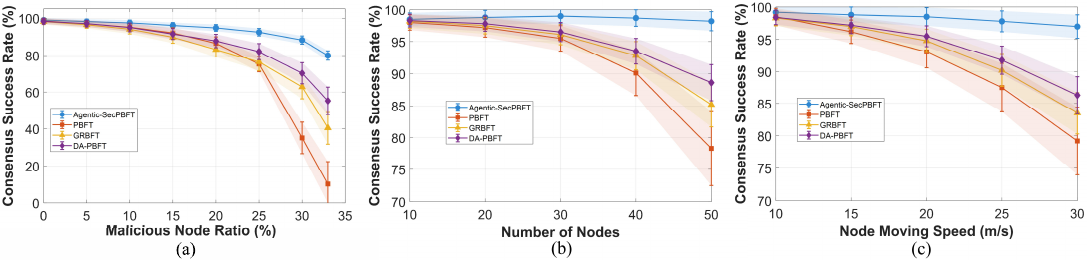}
   \caption{Consensus success rate comparison. (a) Under different malicious node ratios; (b) Under different numbers of nodes; (c) Under different node moving speed.}
\label{fig:CSR}
   \vspace{-0.5cm}
\end{figure*}

For the malicious node ratio, Fig. \ref{fig:CSR} (a) shows Agentic-SecPBFT maintains a high success rate, dropping slightly from $99.0\%$ to $80.0\%$. Standard PBFT plummets drastically from $98.5\%$ to only $10.2\%$, and GRBFT and DA-PBFT fall to $40.5\%$ and $55.2\%$. In Fig. \ref{fig:CSR} (b), when the number of nodes scales from $10$ to $50$, Agentic-SecPBFT remains stable with a success rate ranging from $98.5\%$ to $99.0\%$. PBFT’s success rate drops to $78.3\%$ due to excessive communications. As shown in Fig. \ref{fig:CSR} (c), with the increase in node speed from $10$ m/s to $30$ m/s, Agentic-SecPBFT’s success rate only slightly decreases from $99.2\%$ to $97.0\%$. While PBFT experiences a sharp drop from $98.5\%$ to $79.2\%$, and GRBFT and DA-PBFT also show obvious declines. The reasons for this superior robustness lie in three core designs of Agentic-SecPBFT. It reduces the semantic gap by using CSI spatial correlation and Rician $K$-factor analysis as auxiliary evidence for Sybil detection, while retaining cryptographic authentication and PBFT quorum validation. The hierarchical MADQN enables proactive identification and isolation of malicious nodes, preventing them from disrupting the consensus process. Its cross-layer perception capability fuses SINR and CSI features to accurately distinguish benign node mobility from malicious behavior, avoiding unnecessary view changes that damage consensus stability.

 \begin{figure*}[!t]
\centering
 \includegraphics[width=6.5 in]{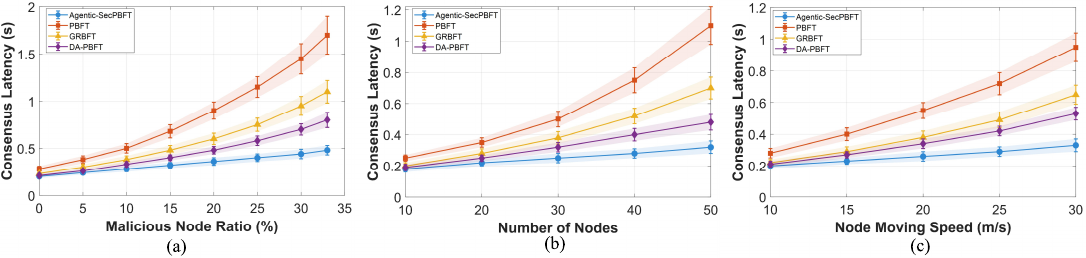}
   \caption{Consensus latency comparison. (a) Under different malicious node ratios; (b) Under different numbers of nodes; (c) Under different node moving speed.}
\label{fig:LA}
   \vspace{-0.5cm}
\end{figure*}

 \begin{figure*}[!t]
\centering
 \includegraphics[width=6.5 in]{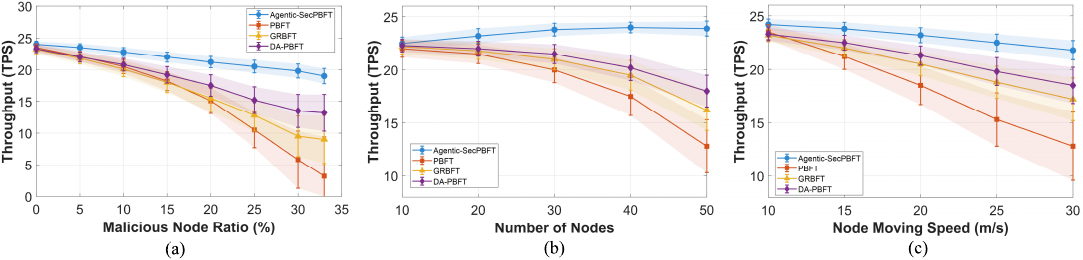}
   \caption{Consensus throughput comparison. (a) Under different malicious node ratios; (b) Under different numbers of nodes; (c) Under different node moving speed.}
\label{fig:TPS}
   \vspace{-0.5cm}
\end{figure*}

As shown in Fig. \ref{fig:LA} (a), when the malicious node ratio increases from $0\%$ to $33\%$, Agentic-SecPBFT’s consensus latency increases moderately from $0.20$ s to $0.48$ s. While PBFT’s latency surges by $507\%$ from $0.28$ s to $1.70$ s due to frequent view changes triggered by malicious nodes. GRBFT and DA-PBFT’s latency reach $1.10$ s and $0.80$ s respectively at $33\%$ malicious nodes. In Fig. \ref{fig:LA} (b), as the number of nodes scales from $10$ to $50$, Agentic-SecPBFT’s latency rises gently from $0.1$8 s to $0.32$ s, whereas PBFT’s latency skyrockets to $1.10$ s due to the exponential growth of redundant communication. With the node moving speed increasing from $10$ m/s to $30$ m/s, as shown in Fig. \ref{fig:LA} (c), Agentic-SecPBFT’s latency only increases from $0.20$ s to $0.33$ s, significantly outperforming the three comparison schemes. Because they will wrongly interpret the connectivity disruptions caused by mobility as malicious silences and trigger unnecessary view changes.
 Agentic-SecPBFT’s low latency is mainly due to the following aspects. The reward function imposes a large negative penalty on unnecessary view changes, incentivizing the model to minimize such latency-consuming operations. Also, LCAs can perform real-time threat flagging and non-critical traffic deprioritization, while permanent removal remains subject to a PBFT-approved membership update. Moreover, its cross-layer awareness capability accurately distinguishes between connectivity issues caused by node mobility and malicious silence, avoiding view changes triggered by misjudgment.

Under different malicious node ratios, Fig. \ref{fig:TPS} (a) shows that Agentic-SecPBFT maintains stable throughput, dropping slightly from $24.1$ TPS to $18.9$ TPS. While PBFT’s throughput collapses to only $3.2$ TPS, and GRBFT and DA-PBFT drop to $9.0$ TPS and $13.2$ TPS. As shown in Fig. \ref{fig:TPS} (b), when the number of nodes increases, Agentic-SecPBFT’s throughput first rises and then stabilizes, peaking at $23.9$ TPS. It will be thanks to COA-driven load balancing that fully utilizes network resources, and it has significantly outperformed the comparison scheme. As shown in Fig. \ref{fig:TPS} (c), with the node moving speed increasing from $10$ m/s to $30$ m/s, Agentic-SecPBFT still retains a high throughput of $21.8$ TPS at 30 m/s, and also better than the other three. 
The stable and high throughput of Agentic-SecPBFT stems from its effective protection of network liveness. The hierarchical MADQN algorithm optimizes the composition of the consensus group, enabling honest nodes to conduct efficient consensus without being overwhelmed by malicious nodes. And the COA realizes global load balancing, improving the utilization efficiency of wireless bandwidth in large-scale networks. In addition, its cross-layer perception of CSI/SINR enables adaptive adjustment to dynamic network topologies, ensuring continuous and efficient communication between honest nodes. Moreover, the dynamic reputation mechanism also avoids punishing honest nodes that have poor channel conditions. It retains its effective data processing capabilities during the consensus.

\subsection{Ablation Study}
To isolate the contributions of three core components, we evaluate the complete framework and three variants: without CSI/SINR context (w/o PHY), without dynamic reputation (w/o Rep.), and without COA coordination (w/o COA), under both low mobility \& low interference and high mobility \& high interference scenarios (they represent a relatively stable and extreme communication environment), as summarized in Table~\ref{tab:ablation}.
Removing physical-layer context causes the sharpest performance degradation in both settings. In the benign scenario, CSR falls from 98.2\% to 89.4\%, ADR from 95.0\% to 87.6\%, and FPR rises to 5.9\%. Under high mobility and high interference, this gap widens notably. CSR drops to 78.3\%, ADR to 76.5\%, and FPR surges to 11.2\%. Without CSI/SINR awareness, the system cannot distinguish channel-induced packet loss from malicious dropping, triggering excessive unnecessary view changes and misjudgments that grow worse in dynamic environments.
Removing dynamic reputation leads to moderate performance loss. CSR, ADR, and FPR shift to 93.7\%, 91.3\%, and 3.8\% in the low-interference case, and deteriorate to 86.4\%, 84.7\%, and 6.8\% under harsh conditions. This demonstrates that temporal trust accumulation suppresses transient misclassification from channel fluctuations, and its stabilizing effect becomes more critical as network instability increases.
Removing COA coordination causes the mildest decline. In the benign scenario, CSR remains at 96.1\%, ADR at 92.0\%, and FPR at 2.7\%. Even under high interference, the metrics stay at 89.8\%, 87.1\%, and 4.9\%. Since LCAs retain full local perception and reputation capabilities, they can independently sustain basic defense and consensus. This verifies that the COA acts as a performance enhancement layer rather than a single point of failure.

\begin{table}[!t]
\caption{Ablation Study Under Two Typical Scenarios}
\label{tab:ablation}
\centering
\resizebox{\linewidth}{!}{
\begin{tabular}{lccc}
\hline
\hline
\multicolumn{4}{c}{\textbf{Low Mobility \& Low Interference}} \\
\hline
Variant & CSR (\%) & ADR (\%) & FPR (\%) \\
\hline
Agentic-SecPBFT       & $98.2 \pm 0.3$ & $95.0 \pm 0.4$ & $1.8 \pm 0.2$ \\
w/o PHY               & $89.4 \pm 0.7$ & $87.6 \pm 0.6$ & $5.9 \pm 0.4$ \\
w/o Rep.              & $93.7 \pm 0.5$ & $91.3 \pm 0.5$ & $3.8 \pm 0.3$ \\
w/o COA               & $96.1 \pm 0.4$ & $92.0 \pm 0.5$ & $2.7 \pm 0.2$ \\
\hline
\multicolumn{4}{c}{\textbf{High Mobility \& High Interference}} \\
\hline
Variant & CSR (\%) & ADR (\%) & FPR (\%) \\
\hline
Agentic-SecPBFT       & $92.5 \pm 0.6$ & $90.2 \pm 0.7$ & $3.5 \pm 0.3$ \\
w/o PHY               & $78.3 \pm 1.2$ & $76.5 \pm 1.1$ & $11.2 \pm 0.8$ \\
w/o Rep.              & $86.4 \pm 0.9$ & $84.7 \pm 0.8$ & $6.8 \pm 0.5$ \\
w/o COA               & $89.8 \pm 0.7$ & $87.1 \pm 0.8$ & $4.9 \pm 0.4$ \\
\hline
\hline
\end{tabular}}
\end{table}

\section{Conclusion}\label{sec-vii}

In this paper, we demonstrate that standard PBFT and its static security enhancements are highly susceptible to intelligent, coordinated threats in wireless networks, such as Sybil and collusion attacks. To address these challenges, we propose a security framework, Agentic-SecPBFT, leveraging an Agentic AI-empowered multi-agent system. It employs a distributed intelligent model, including COA and LCA. The LCA detects suspicious behavior by observing local network conditions and recommends guarded security actions in real time. %CSI is treated as probabilistic auxiliary evidence rather than an unforgeable physical identity, and permanent exclusion remains governed by PBFT membership rules. 
The COA supports cross-node learning through authenticated, versioned updates; its failure can reduce adaptation quality but cannot directly violate the PBFT commit quorum.
The results show that this solution achieves substantial improvements in attack resilience, maintaining a high consensus success rate, low latency, and stable throughput even under severe attacks. Meanwhile, the ablation experiments confirmed the positive benefits of each module. This framework is expected to provide a new secure path for wireless blockchain consensus.
\bibliographystyle{IEEEtran}
\bibliography{IEEEabrv,mylib}

\vfill

\end{document}